\documentclass{kapproc} 

\usepackage{procps} 

\usepackage[dvips]{graphicx}

\upperandlowercase

\normallatexbib 

\begin{document}

\articletitle
{Merging processes and SZ effect}

\author{P. M. Koch\altaffilmark{1,2}, Ph. Jetzer\altaffilmark{2}}

\altaffiltext{1}{Institute of Astronomy and Astrophysics, Academia Sinica, P.O.Box 23-141, \\Taipei 106, Taiwan, R.O.C.\\
{\tt \small pmkoch@asiaa.sinica.edu.tw}}
\altaffiltext{2}{University of Zurich, Institute for Theoretical Physics,
Winterthurerstrasse 190,\\ CH-8057 Zurich, Switzerland\\
{\tt \small jetzer@physik.unizh.ch}}



\section{Introduction}
\noindent
Recent observations reveal detailed structures of mergers in 
galaxy clusters. Moving cold fronts could be identified as remnants
of subclusters. They are believed to mark the late stage of a merging process \cite{vi01}.
Dynamics, geometry and extension of the subclusters can be very different: Spherically shaped bodies 
with a radius of $\approx 20\,kpc$ up to $\approx 400\,kpc$ and Mach numbers between
$\mathcal{M}\approx 0.4$ and $\mathcal{M}\approx 1.5$ are observed.\\
In order to analyse the Sunyaev-Zeldovich (SZ) signal,
we calculate the modified pressure profile for an ongoing merger.
For subsonic mergers we present first results of our work in progress.

\section{Subsonic mergers}
\noindent
Compressibility effects 
are still small for $\mathcal{M}\approx 0.4-0.8$. We assume a merger with zero impact parameter
and a merger axis parallel to the line of sight.
The integration 
through the cluster center is divided into the regions ahead and behind 
of the subcluster and the subcluster itself.
Ahead, an excess pressure contribution due to the moving body - in top of the 
hydrostatic pressure - is expected.
For the subcluster itself we adopt data from X-ray observations.\\
We define $\Delta SZ_M=\frac{\Delta I_M(x)}{\Delta I(x)}$ as the ratio of 
the merger influenced SZ signal compared to the standard thermal SZ effect. Not expecting non-thermal
contributions for an incompressible merger, there is no frequency ($x$) dependence and $\Delta I_M$ is
given by the sum of the integrated pressure contributions from the subregions. Assuming only small 
turbulences in the region behind, $\Delta SZ_M$ becomes:
\begin{eqnarray}
\Delta SZ_M=1 & + & \frac{1}{2\int_0^{r_l}f(r)\,dr}\left[\int_{r_{M,1}}^{r_l}f(r)\frac{1}{2}\mathcal{M}_{\infty}^2
            (1-\frac{v(r)^2}{u_{\infty}^2})\,dr   \right.      \nonumber  \\
            & - & \left.\int_{r_{M,1}}^{r_{M,2}}f(r)\,dr+a\frac{n_{e,M}T_M}{n_{e,0}T}\right], \nonumber
\end{eqnarray}
where $f(r)$ is the cluster gas profile. $r_{M,1}$ and $r_{M,2}$ define the boundaries of the subcluster
with extension $a$.
$r_l$ is the cluster limiting radius. $n_{e,M}$ and $T_M$ are the subcluster central density and temperature,
respectively. $u_{\infty}$ is the free streaming velocity and $v(r)$ is the velocity profile in front of the subcluster.\\
The first term in square brackets (excess pressure) always gives a positive contribution. The
third (subcluster) must be compared to the negative second  accounting for the removed gas. The following 
two figures
illustrate the excess pressure contribution for selected parameters and subcluster geometries. 

\begin{center}
 \begin{minipage}[b]{.49\linewidth}
  \includegraphics[width=\linewidth]{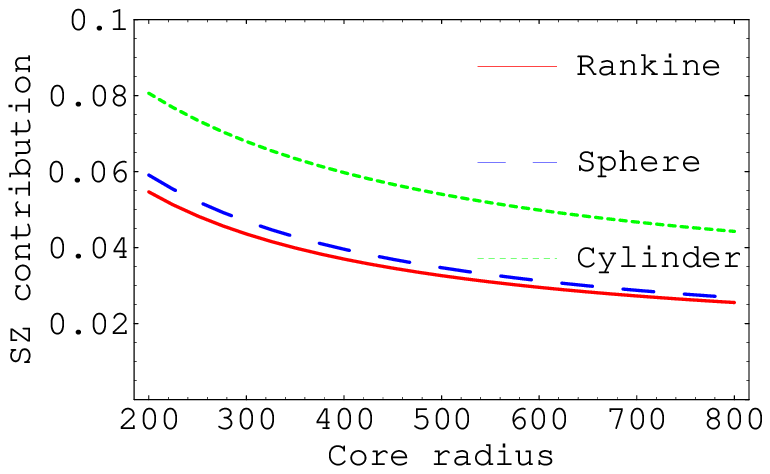}
  \footnotesize
  \centering $a=200\,kpc$, forefront at $r=0$.
 \end{minipage}
 \hfill
  \begin{minipage}[b]{.49\linewidth}
  \includegraphics[width=\linewidth]{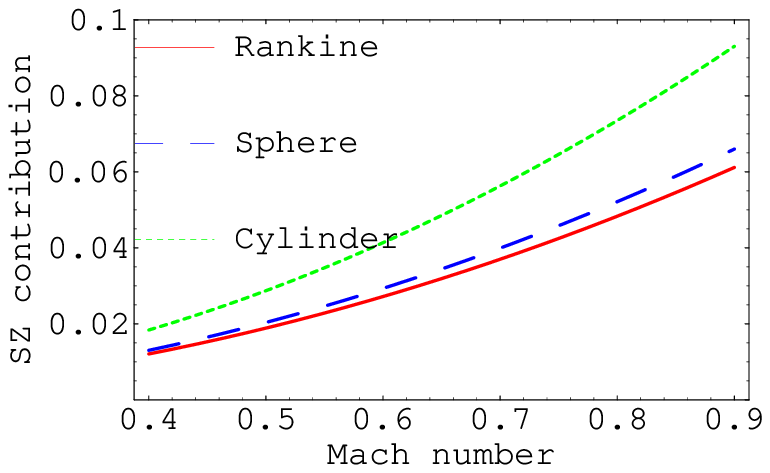}
 \footnotesize
  \centering $a=200\,kpc$, forefront at $r=0$.
 \end{minipage}
\end{center}

\noindent
The contribution from the subcluster versus the removed gas is shown in the next two figures for $\frac{n_{e,M}T_M}{n_{e,0}T}\approx 2$.
$l$ is the location of the subcluster center.

\begin{center}
 \begin{minipage}[b]{.49\linewidth}
  \includegraphics[width=\linewidth]{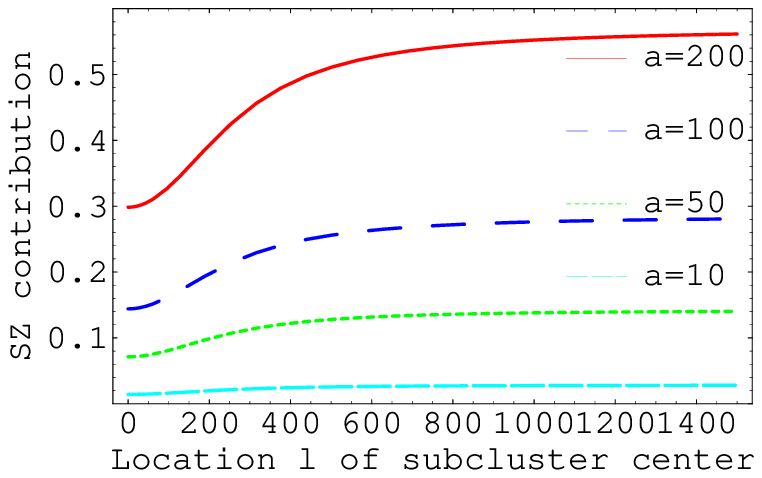}
  \footnotesize
  \centering Standard cluster: $r_c=250\,kpc$, $r_l=1500\,kpc$.
 \end{minipage}
 \hfill
  \begin{minipage}[b]{.49\linewidth}
  \includegraphics[width=\linewidth]{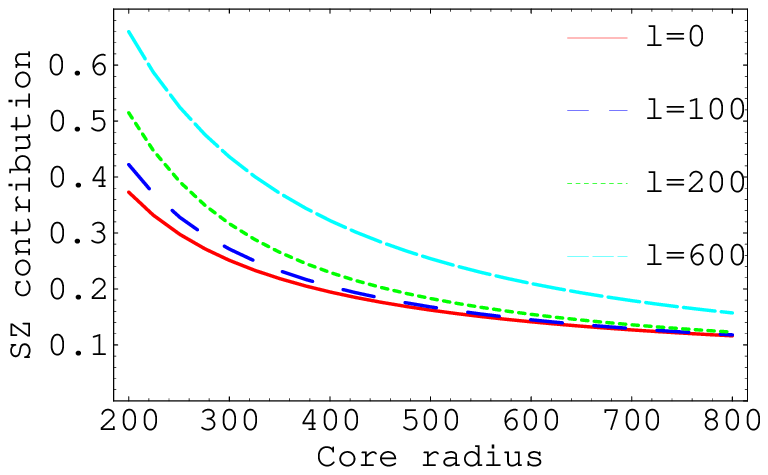}
 \footnotesize
  \centering Standard cluster: $r_l=1500\,kpc$, $a=200\,kpc$.
 \end{minipage}
\end{center}

\section{Conclusion}
\noindent  
Beside magnetic fields \cite{ko03} and large cooling flows \cite{ko02}, also mergers can produce an enhanced 
standard SZ signal. 
Transonic mergers ($\mathcal{M}>1$) with shock heated regions can vary 
both intensity and spectral shape of the SZ effect.

\pagebreak


\begin{chapthebibliography}{1}

\bibitem{vi01} Vikhlinin, A., Markevitch, M., Murray, S.M., 2001. ApJ. 551, 160
\bibitem{ko03} Koch, P.M., Jetzer, Ph., Puy, D., 2003. New Astronomy 8, 1
\bibitem{ko02} Koch, P.M., Jetzer, Ph., Puy, D., 2002. New Astronomy 7, 587


\end{chapthebibliography}
\end{document}